# Dynamical Correlation Functions and Finite-size Scaling in Ruijsenaars-Schneider Model

Hitoshi Konno[*]

*Yukawa Institute for Theoretical Physics,
Kyoto University, Kyoto 606-01, Japan*[†]

ABSTRACT

The trigonometric Ruijsenaars-Schneider model is diagonalized by means of the Macdonald symmetric functions. We evaluate the dynamical density-density correlation function and the one-particle retarded Green function as well as their thermodynamic limit. Based on these results and finite-size scaling analysis, we show that the low-energy behavior of the model is described by the $C = 1$ Gaussian conformal field theory under a new fractional selection rule for the quantum numbers labeling the critical exponents.

PACS number: 03.65.-w, 05.30.-d, 02.90.+p, 12.90.+b

[*] Yukawa Fellow
[†] konno@yukawa.kyoto-u.ac.jp

# 1   Introduction

In the recent study of the Calogero-Sutherland model (CSM)[1, 2] Yangian symmetry and Jack symmetric functions have played a central role[3~10].

Yangians are quantum groups associated with the rational solution of the Yang-Baxter equation (YBE)[11]. Realizing the Yangian $Y(gl_2)$ in terms of the degenerate affine Hecke algebra, Bernard et al. have formulated the spin generalization of the CSM[4]. This allows us to understand the integrability of the CSM based on the YBE analogously to other integrable models such as Heisenberg spin chains and to construct the space of states based on the representation of the Yangian[12].

The Jack symmetric function is a multi-variable orthogonal symmetric polynomial. It has been shown that the space of states of the CSM is spanned by the Jack symmetric functions[8, 7, 9, 10]. Furthermore, the orthogonality property of the functions allows one to calculate the dynamical correlation functions[9, 10]. Remarkably, the resultant density-density correlation function turns out to provide a certain universal correlation function in the random matrix theory at the special coupling constants $g = 1/2, 1$ and 2[13].

The CSM has its one-parameter deformation known as the trigonometric Ruijsenaars-Schneider model (RSM)[14]. In the recent paper[15], we have considered this model and shown that its spin generalization possesses the quantum affine symmetry generated by the level-0 action of $U_q(\widehat{gl}_2)$[16]. The algebra $U_q(\widehat{gl}_2)$ at level-0 is known as a trigonometric extension of the Yangian $Y(gl_2)$. Hence one may expect a similar construction of the space of states based on the representation of $U_q(\widehat{gl}_2)$ at the level-0 to the CSM done by the Yangian representation. We also have shown that the (spin-less) trigonometric RSM is diagonalized by using the Macdonald symmetric functions[17], in the same way as the CSM is by the Jack symmetric functions. The Macdonald symmetric function is known to be a one-parameter deformation of the Jack symmetric function.

The RSM has been discussed by many people in connection to wide variety of fields such as sine-Gordon theory[14, 18, 19], Toda theory[14, 20, 21], G/G gauged Wess-Zumino-Witten (WZW) model[22] as well as 4d supersymmetric Yang-Mills theory[23, 24].

In this paper, we consider the calculation of dynamical correlation functions of the trigonometric RSM. In [15], we have briefly reported the results on the dynamical density-density correlation function and the one-particle retarded Green function in the finite system with particle number $N$ and circumference $L$, on which particles are confined. We present here the details of the calculation. Our method is based on the Macdonald symmetric functions and is completely parallel to the one done in the CSM based on the Jack symmetric functions[9, 10]. For this purpose, we prepare some formulae of the Macdonald symmetric functions in section 3. We next investigate the thermodynamic limit, $N, L \to \infty$ with $N/L$ fixed, of the dynamical correlation functions and obtain natural one-parameter deformed expressions of those of the CSM. Physical significance of these results is not so clear at this stage. However the argument by Gorsky and Nekrasov[22], which shows the equivalence between the trigonometric RSM and the G/G gauged WZW model with a certain Wilson line insertion suggests that our results should be connected with some amplitudes in the latter model.

We also investigate the statistics of the excitations of the model. We show that for a rational coupling constant $g$, they obey the fractional exclusion statistics à la Haldane with statistical parameter $g$[25]. We stress the role of the dual description of the states associated with a conjugate partition $\lambda'$ and a coupling constant $1/g$. Its consistency with a description associated with a partition $\lambda$ and the coupling constant $g$ selects the physical processes.

On the other hand, it has been shown by Kawakami and Yang that the low energy behavior of the CSM is described by the $C = 1$ Gaussian CFT[26]. However the critical exponents obtained by Ha[10] based on his result of the one-particle Green function seem not to agree with those in [26]. This is due to the fact that Kawakami and Yang considered the correlation functions of operators associated with real fermions or bosons, whereas Ha treated those associated with the pseudo-particles. Although Ha has explained his results based on the low-energy effective theory of the pseudo-particles which indicates the fractional exchange statistics, their connection to the standard CFT argument is still unknown. Here we settle this point. For this purpose, we derive a new selection rule (6.9) for pseudo-particles obeying the generalized Pauli exclusion principle, which restrict the possible quantum numbers labeling the critical exponents. Under this rule, we show that the critical exponents of the Green function both in the CSM and in the trigonometric RSM agree with the CFT predictions. We thus conclude that the trigonometric RSM belongs to the same universality class as the CSM, i.e. the $C = 1$ Gaussian theory with the compactification radius $1/\sqrt{g}$. As discussed in [15], this indicates that the model behaves as a Tomonaga-Luttinger liquid[27, 26] in the low energy region.

This paper is organized as follows. In section 2, we review the trigono-metric RSM and fix the notations. In section 3, we give a brief review of the theory of Macdonald symmetric functions and present some useful formulae. In section 4, we discuss the diagonalization of the trigonometric RSM and investigate the statistics of the excitations for a rational coupling constant $g$. We show that the excitations of the model obey the fractional exclusion statistics with statistical parameter $g$. Section 5 is devoted to a detailed description of the calculation of the dynamical correlation functions. The thermodynamic limits of the results and their low-energy asymptotic behavior are derived. In section 6, we consider the finite-size scaling corrections to the excitation energy and momentum spectrum and identify the CFT describing the low energy behavior of the model. We also derive a new selection rule of the quantum numbers.

## 2  The trigonometric Ruijsenaars-Schneider model

The trigonometric Ruijsenaars-Schneider model (TRSM) is an integrable quantum mechanical system of $N-$relativistic particles[14]. Let $\theta_j$ ($j = 1, 2, .., N$) be the rapidity variables and $x_j$ be their canonically conjugate variables. We impose the canonical commutation relations

$$[x_j, \theta_l] = i\delta_{j,l}, \tag{2.1}$$

with $\hbar = 1$. We use the representation $\theta_l = -i\partial/\partial x_l$. The model is described by the following Hamiltonian $H$ and momentum operator $P$

$$H = \frac{c^2}{2}(H_{-1} + H_1), \qquad P = \frac{c}{2}(H_{-1} - H_1) \tag{2.2}$$

as well as the following $N-$independent integrals of motion $H_k$ (or $H_{-k}$) ($k = 1, 2, .., N$)

$$H_{\pm k} = \sum_{\substack{I \subset \{1,2,..,N\} \\ |I|=k}} \prod_{\substack{i \in I \\ j \notin I}} h^{\pm}(x_i - x_j)^{1/2} e^{-1/c \sum_{i \in I} \theta_i} \prod_{\substack{l \in I \\ m \notin I}} h^{\pm}(x_m - x_l)^{1/2} \tag{2.3}$$

$$h^{\pm}(x_i - x_j) = \frac{\sin\frac{\pi}{L}(x_i - x_j \mp ig/c)}{\sin\frac{\pi}{L}(x_i - x_j)}. \tag{2.4}$$

Here $c$ being the speed of light, $L \in \mathbf{R}_{>0}$ and $g \in \mathbf{Q}$ being the coupling constant. The particles are confined on the ring of length $L$ in the space of $\{x_j\}$. We normalize the mass equal to one. Notice that $H_{\pm k}$ ($k = 1, 2, ..., N$) commute with each other

$$[H_k, H_l] = 0 \qquad \forall k, l \in \{-N, -N+1, ..., N\}. \tag{2.5}$$

The Lorentz boost generator $B$ is given by

$$B = -\frac{1}{c}\sum_{i=1}^{N} x_i. \tag{2.6}$$

The model possesses Poincaré invariance in the sense that the operators $H$, $P$, and $B$ satisfy the Poincaré algebra[1]:

$$[H,P]=0, \qquad [H,B]=iP, \qquad [P,B]=i\frac{H}{c^2}. \tag{2.7}$$

In the 'non-relativistic' limit $c \to \infty$, we recover the Hamiltonian of CSM

$$\lim(H - Nc^2)$$
$$= -\sum_{j=1}^{N}\frac{1}{2}\Big(\frac{\partial}{\partial x_j}\Big)^2 + g(g-1)\sum_{1\leq j<k\leq N}\frac{(\pi/L)^2}{\sin^2\frac{\pi}{L}(x_j-x_k)} + \text{const.} \tag{2.8}$$

It is also known that the integrals of motions $H_k$ has a deep connection with the Macdonald operators. The Macdonald operators $D_k(p,t)$ ($k = 1,2,..,N$) are defined by[17]

$$D_k(p,t) = t^{k(k-1)/2} \sum_{\substack{I \subset \{1,2,..,N\} \\ |I|=k}} \prod_{\substack{i \in I \\ j \notin I}} \frac{tz_i - z_j}{z_i - z_j} \prod_{i \in I} p^{\vartheta_i} \tag{2.9}$$

for $N$-variables $z_j$ ($j = 1,2,..,N$) with $t,p \in \mathbf{C}^{\times}$. $p^{\vartheta_j}$ denotes the shift operator $(p^{\vartheta_j}f)(z_1,...,z_N) = f(z_1,...,pz_j,...,z_N)$. The operators $D_k(p,t)$ ($k = 1,2,..,N$) can be simultaneously diagonalized by the Macdonald symmetric functions (see §3).

In order to show the connection between $D_k$ and $H_k$, let us set

$$p = e^{-2\pi/Lc}, \quad t = p^g \tag{2.10}$$

$$z_j = e^{i\frac{2\pi}{L}x_j}, \quad p^{\pm\vartheta_j} = e^{\mp\frac{2\pi}{Lc}z_j\frac{\partial}{\partial z_j}}. \tag{2.11}$$

Then, by using the function

$$\Delta_N = \prod_{\substack{j,k=1 \\ j\neq k}}^{N} \frac{(z_j/z_k;p)_\infty}{(tz_j/z_k;p)_\infty}, \tag{2.12}$$

---

[1]One should note that the Poincaré invariance does not guarantee that the Hamiltonian (2.2) defines an interacting system of relativistic particles satisfying the causality. In this paper, we use the terminology 'relativistic' only in the kinematical sense.

where
$$(x;p)_\infty = \prod_{n=0}^{\infty}(1-xp^n), \qquad (2.13)$$

one has [28]

$$\Delta_N^{-1/2} H_{\pm k} \Delta_N^{1/2} = t^{\mp k(N-1)/2} D_k(p^{\pm 1}, t^{\pm 1}) \qquad k=1,2,..,N. \qquad (2.14)$$

## 3 Macdonald symmetric functions

In this section we briefly review the Macdonald symmetric functions[17]. Let $\lambda = (\lambda_1, \cdots, \lambda_N)$, $\lambda_1 \geq \cdots \geq \lambda_N \geq 0$, $\lambda_j \in \mathbf{Z}$ be a partition. The Macdonald symmetric function $P_\lambda(z;p,t) \equiv P_\lambda(z_1,..,z_N;p,t)$ is an eigen function of the Macdonald operators,

$$D_k(p^{\pm 1}, t^{\pm 1}) P_\lambda(z;p,t) = \Big(\sum_{i_1 < \cdots < i_k} \prod_{l=1}^{k} t^{N-i_l} p^{\lambda_{i_l}}\Big) P_\lambda(z;p,t). \qquad (3.1)$$

The Macdonald symmetric functions form an orthogonal basis of a ring of symmetric polynomials $\Lambda_N$ in the $N$ variables $z = (z_1, z_2, .., z_N)$ with coefficients in $\mathbf{Q}(p,t)$. Let $l(\lambda)$ be the length of $\lambda$, i.e. $l(\lambda) = \max\{j|\lambda_j > 0\}$. To each partition $\lambda$, we assign a Young diagram $\mathcal{D}(\lambda) = \{(i,j)|1 \leq i \leq l(\lambda), 1 \leq j \leq \lambda_i, \quad i,j \in \mathbf{Z}_{>0}\}$. We denote the conjugate partition of $\lambda$ by $\lambda'$ corresponding to the transpose of the Young diagram $\mathcal{D}(\lambda)$. For each cell $\gamma = (i,j)$ of $\mathcal{D}(\lambda)$, we define the quantities $a(\gamma) = \lambda_i - j$, $a'(\gamma) = j-1$, $l(\gamma) = \lambda'_j - i$ and $l'(\gamma) = i - 1$.

Define the scaler product in $\Lambda_N$ by

$$\begin{aligned}
<f,g>'_{p,t;N} &= \frac{1}{n!}\oint \prod_{j=1}^{N} \frac{dz_j}{2\pi i z_j} \bar{f}(z) g(z) \Delta_N \\
&= \frac{1}{n!}(\text{constant term in } \overline{f(z)} g(z) \Delta_N), \qquad (3.2)
\end{aligned}$$

with $\overline{f(z)} = f(1/z_1, .., 1/z_N)$, then the orthogonality of the two Macdonald symmetric functions is stated as follows.

$$<P_\mu, P_\lambda>'_{p,t} = \delta_{\mu,\lambda} \, A_N \, \frac{h_{\lambda'}(t,p)}{h_\lambda(p,t)} \, \mathcal{N}(\lambda), \qquad (3.3)$$

where

$$A_N = <1,1>'_{p,t;N} = \prod_{j=1}^{N} \frac{(pt^{j-1};p)_\infty (t;p)_\infty}{(t^j;p)_\infty (p;p)_\infty}, \qquad (3.4)$$

$$h_\lambda(p,t) = \prod_{\gamma \in \lambda} \left(1 - p^{a(\gamma)} t^{l(\gamma)+1}\right), \qquad (3.5)$$

$$h_{\lambda'}(t,p) = \prod_{\gamma \in \lambda} \left(1 - p^{a(\gamma)+1} t^{l(\gamma)}\right), \qquad (3.6)$$

$$\mathcal{N}(\lambda) = \prod_{\gamma \in \lambda} \frac{1 - p^{a'(\gamma)} t^{N-l'(\gamma)}}{1 - p^{a'(\gamma)+1} t^{N-l'(\gamma)-1}}. \qquad (3.7)$$

The $P_\lambda(z;p,t)$ satisfies the following duality property:

$$\sum_\lambda P_\lambda(z;p,t) P_{\lambda'}(w;t,p) = \prod_{i,j}(1 + z_i w_j). \qquad (3.8)$$

In the calculation of dynamical correlation functions (§5), the following expansion formulae[17, 29] are essential.

$$\sum_{j=1}^{N} z_j^m = (1 - p^m) \sum_{\substack{\lambda \\ |\lambda|=m}} \frac{\chi^\lambda(p,t)}{h_{\lambda'}(t,p)} P_\lambda(z;p,t), \qquad (3.9)$$

$$\prod_{j=1}^{N} \frac{(z_j;p)_\infty}{(az_j;p)_\infty} = \sum_\lambda \frac{a^{|\lambda|}(a^{-1})_\lambda^{(p,t)}}{h_{\lambda'}(t,p)} P_\lambda(z_j;p,t) \qquad (3.10)$$

where $a \in \mathbf{C}^\times$ and

$$\chi^\lambda(p,t) = \prod_{\substack{\gamma \in \lambda \\ \gamma \neq (1,1)}} \left(t^{l'(\gamma)} - p^{a'(\gamma)}\right), \qquad (3.11)$$

$$(a)_\lambda^{(p,t)} = \prod_{\gamma \in \lambda} \left(t^{l'(\gamma)} - p^{a'(\gamma)} a\right). \qquad (3.12)$$

## 4 Diagonalization

We here consider the diagonalization of the TRSM and investigate the statistics of the elementary excitations.

From (2.14) and (3.1), one can diagonalize $H$ and $P$ by the functions

$$\Phi_{N;\lambda}(z) = \Delta_N^{1/2} P_\lambda(z;p,t) \qquad (4.1)$$

with the following eigenvalues

$$E_N(\lambda) = c^2 \sum_{j=1}^{N} \text{ch} \frac{\theta_j}{c}, \qquad P_N(\lambda) = c \sum_{j=1}^{N} \text{sh} \frac{\theta_j}{c}, \qquad (4.2)$$

where

$$\theta_j = \frac{2\pi}{L} \left\{ \lambda_j + g \left( \frac{N+1}{2} - j \right) \right\}. \qquad (4.3)$$

The model thus can be regarded as an ideal gas of $N$-relativistic pseudo-particles with the pseudo-rapidities (4.3).

The formula (4.3) implies the minimum difference between two consecutive pseudo-rapidities

$$\theta_i - \theta_{i+1} \geq \frac{2\pi}{L} g. \qquad (4.4)$$

When $g = 0$ there are no restrictions, and more than two pseudo-particles can occupy the same quantum sate so that the statistics is bosonic. In the case $g = 1$, a pseudo-particle occupy one of the quantized states by the unit $2\pi/L$ obeying Pauli exclusion principle so that the statistics is fermionic. For generic $g$, the pseudo-particles can be said to obey the generalized Pauli exclusion principle[25].

The ground state is given by the function $\Psi_\emptyset(z) = \Delta_N^{1/2}$ corresponding to the empty partition $\lambda = \emptyset$. Note $P_\emptyset(z; p, t) = 1$. The ground state energy and momentum eigenvalues are obtained as

$$E_N^{(0)} = c^2 \text{sh} \frac{\pi g N}{Lc} \Big/ \text{sh} \frac{\pi g}{Lc} \qquad (4.5)$$

and $P_N^{(0)} = 0$, respectively. One should note that the ground state can be described as a filled Fermi sea with pseudo-momenta $P_j^{(0)} = c \, \text{sh} \frac{\theta_j}{c}$ with $-\theta_F \leq \theta_j \leq \theta_F$ $j = 1, 2, .., N$, where

$$\theta_F = \frac{\pi g (N-1)}{L}. \qquad (4.6)$$

Let us next consider the excitations. We consider the case with the rational coupling $g = r/s$ with $r$ and $s$ ($\neq 0$) being coprime integers.

In order to describe the excitations, we assign a motif to each set of pseudo-rapidities $\{\theta_j\}$ $j = 1, 2, .., N$. A motif is a sequence of numbers 0 or 1 with totally $N$ of 1s. It can be constructed as follows[10]. Consider a one-dimensional lattice whose lattice spacing is $2\pi/sL$, and assign each lattice point with 1 if that point coincides with the value of one of the pseudo-rapidities, otherwise assign 0. The allowed number of 0s between each pair of 1s is $r - 1 + ns$ with $n \in \mathbf{Z}_{\geq 0}$. Among the $r - 1 + ns$ 0s, we regard $r - 1$ consecutive 0s as to be bounded to each 1 and the other 0s as unbounded.

We define a quasi-particle as 1 which lattice point $\theta$ satisfies $\theta > \theta_F$ or $\theta < -\theta_F$. Out of $r - 1 + ns$ 0s, we define a quasi-hole as $s$ consecutive un-bounded 0s whose lattice points are lying in the region $[-\theta_F, \theta_F]$.

By using the duality property (3.8), one has a dual description of the excited states, where the roles of quasi-particles and -holes are exchanged. More precisely, let $\tilde{p} = t, \tilde{t} = p$. Then $\tilde{t} = \tilde{p}^{1/g}$ and $D_k(\tilde{p}, \tilde{t}) = D_k(t, p)$. The operator $D_k(\tilde{p}, \tilde{t})$ are diagonalized by the dual Macdonald polynomials $P_{\lambda'}(t, p)$ yielding the eigenvalues $\sum_{j=1}^{N} \prod_{l=1}^{k} p^{N-i_l} t^{\lambda'_l}$. Hence if one defines the dual integrals of motion $\tilde{H}_{\pm k}$ by

$$\tilde{H}_{\pm k} = \tilde{t}^{\mp k(N-1)/2} \tilde{\Delta}_N^{1/2} D_k(\tilde{p}, \tilde{t}) \tilde{\Delta}_N^{-1/2} \tag{4.7}$$

with $\tilde{\Delta}_N = \Delta_N|_{p \leftrightarrow t}$, then for $\tilde{H} = \frac{\tilde{c}^2}{2}(\tilde{H}_{-1} + \tilde{H}_1)$ one has

$$\tilde{H} \tilde{\Delta}_N^{1/2} P_{\lambda'}(\tilde{p}, \tilde{t}) = \left( g^2 \tilde{c}^2 \sum_{j=1}^{N} \text{ch} \frac{\tilde{\vartheta}_j}{\tilde{c}} \right) \tilde{\Delta}_N^{1/2} P_{\lambda'}(\tilde{p}, \tilde{t}) \tag{4.8}$$

with $\tilde{c} = c/g$ and $\tilde{\vartheta}_j = \frac{2\pi}{L}(\lambda'_j + \frac{1}{g}(\frac{N+1}{2} - j))$.

For the dual rapidities $\{\tilde{\vartheta}_j\}$, we define a dual motif in the same way as in the above except for changing $r$ and $s$. In the dual picture, the quasi-hole is defined as 1 which lattice point $\tilde{\vartheta}$ satisfies $\tilde{\vartheta} > \tilde{\vartheta}_F$ or $\tilde{\vartheta} < -\tilde{\vartheta}_F$ and the quasi-particle is defined as $r$ consecutive un-bounded 0s lying in $[-\tilde{\vartheta}_F, \tilde{\vartheta}_F]$ with $\tilde{\vartheta}_F = \frac{\pi(N-1)}{gL}$.

We require that a physical state should be described by the above two descriptions consistently. Namely, for a given $\lambda$, we regard a corresponding state as physical only if the number of allowed quasi-particles and -holes obtained in the one description coincides with those obtained in the dual description. For example, in the case $g > 1$, the partition $\lambda = (\underbrace{r-1, .., r-1}_{s})$ seems to be able to give excitations of $r - 1$ quasi-particles and $s$ quasi-holes. However in the dual picture, this case is forbidden because of the condition that the un-bounded 0s corresponding to quasi-particles should be in $[-\tilde{\vartheta}_F, \tilde{\vartheta}_F]$. In such case, we regard the excitations are forbidden.

Now require that the number of quasi-holes should be an integer. Then from the above requirement and (4.3) one can find that for quasi-holes exci-tation, there is the minimum number, which is equal to $r$, and it is accompanied by $s$ quasi-particles excitation. The minimal partition corresponding to this excitation is given by $(\underbrace{r,..,r}_{s})$. This indicates that the quasi-particles and -holes obey the fractional exclusion statistics with the statistical parameter $g = r/s$ as in the CSM[25, 10]. We will show later (§5) that the resultant physical state under this requirement is consistent to the description of the intermediate state both in the density-density correlation function and one-particle Green function.

If one allows the change of the total particle number $N$, there is another way of obtaining excitations. Namely, quasi-holes can be excited by destroying some of the pseudo-particles in the ground state configuration. As discussed by Ha in the CSM[10], this case also leads to the same fractional exclusion statistics as in the above. Namely, for integer number of quasi-holes excitation, the quasi-hole should be excited by the unit consisting of $r$ quasi-holes. The one unit excitation is then possible by the $s$ pseudo-particles destruction from the ground state.

In general, allowing the change of the total particle number, excitations are obtained by mixing the above two types of excitations. A typical example is the intermediate state of the one-particle Green function, where one pseudo-particle destruction from the ground state configuration should be taken into account (see §5).

One should also note that the pseudo-rapidities (4.3) obey the following Bethe ansatz like equations:

$$L\theta_j = 2\pi I_j + \pi(g-1)\sum_{l=1}^{N} \mathrm{sgn}(\theta_j - \theta_l), \tag{4.9}$$

with $I_j = \lambda_j + \frac{N+1}{2} - j$. In the thermodynamic limit, (4.9) yields the relation

$$1 = g\rho_P(\theta) + \rho_H(\theta), \tag{4.10}$$

where $\rho_P(\theta)$ (resp.$\rho_H(\theta)$) are the particle (resp. hole) density. This also indicates the fractional statistics, i.e. one unit increase of the density of the particles of rapidity $\theta$ is accompanied by decrease of the hole density by $g$-unites[30].

## 5 Dynamical correlation functions

The orthogonality property of the Macdonald symmetric functions allows us to calculate the dynamical density-density correlation function $\langle 0|\rho(\xi,\tau)\rho(0,0)|0\rangle$ as well as the one-particle retarded Green function $\langle 0|\Psi^\dagger(\xi,\tau)\Psi(0,0)|0\rangle$, where $\xi$ and $\tau$ are real space and time coordinates, respectively.

## 5.1 Finite system

The dynamical density-density correlation function is defined by

$$\langle 0|\rho(\xi,\tau)\rho(0,0)|0\rangle = \frac{{}_N\langle 0|\rho(\xi,\tau)\rho(0,0)|0\rangle_N}{{}_N\langle 0|0\rangle_N} \tag{5.1}$$

Here the density operator $\rho(\xi,\tau)$ is defined by

$$\rho(\xi,\tau) = e^{i(H\tau-P\xi)}\rho(0,0)e^{-i(H\tau-P\xi)},$$
$$\rho(0,0) = \sum_{j=1}^{N}\delta(x_j) - N/L, \tag{5.2}$$

and $|0\rangle_N$ denotes the $N-$particle ground state, i.e. $|0\rangle_N = \Delta_N^{1/2}$. Combining (3.9) with the formula

$$\rho(0,0) = \frac{1}{L}\sum_{m=1}^{\infty}\sum_{j=1}^{N}(z_j^{-m} + z_j^m), \tag{5.3}$$

one can expand the state $\rho(0,0)|0\rangle_N$ in terms of the $N-$particle eigen states $\Phi_{N;\lambda}(z)$ (4.1). Then using the orthogonality (3.3), we obtain the density-density correlation function as follows[15].

$$\langle 0|\rho(\xi,\tau)\rho(0,0)|0\rangle$$
$$= \frac{2}{L^2}\sum_{\lambda}\frac{(1-p^{|\lambda|})^2(\chi^\lambda(p,t))^2}{h_\lambda(p,t)h_{\lambda'}(t,p)}\mathcal{N}(\lambda)\cos(P_N(\lambda)\xi)e^{-i\mathcal{E}_N(\lambda)\tau}, \tag{5.4}$$

where $\mathcal{E}_N(\lambda) = E_N(\lambda) - E_N^{(0)}$ and $P_N(\lambda)$ are the excitation energy and momentum, respectively.

One should remark that the factor $\chi^\lambda(p,t)$ in (5.4) vanishes if the Young diagram $\mathcal{D}(\lambda)$ contains the lattice point $(s+1, r+1)$. Because of the arguments in §4, this indicates that only the excited states which can contribute to the intermediate state are those consist of minimal $s-$quasi-particle and $r-$quasi-hole excitations. As in the CSM, this supports the fact that the quasi-particles and the quasi-holes obey the fractional exclusion statistics with statistical parameter $g = r/s$.

The one-particle retarded Green function is defined by

$$\langle 0|\Psi^\dagger(\xi,\tau)\Psi(0,0)|0\rangle = \frac{{}_{N+1}\langle 0|\Psi^\dagger(\xi,\tau)\Psi(0,0)|0\rangle_{N+1}}{{}_{N+1}\langle 0|0\rangle_{N+1}}. \quad (5.5)$$

Here $\Psi^\dagger$ and $\Psi$ are the creation and annihilation operators of the pseudo-particles, respectively, and

$$\Psi^\dagger(\xi,\tau) = e^{i(H\tau-P\xi)}\Psi^\dagger(0,0)e^{-i(H\tau-P\xi)}. \quad (5.6)$$

The action of $\Psi(0,0)$ on the ground state $|0\rangle_{N+1}$ is defined by

$$\begin{aligned}\Psi(0,0)|0\rangle_{N+1} &= \Delta_N^{1/2}\Big|_{z_{N+1}=1} \\ &= \prod_{j=1}^N \Big(\frac{(1/z_j;p)_\infty(z_j;p)_\infty}{(t/z_j;p)_\infty(tz_j;p)_\infty}\Big)^{1/2}\Delta_N^{1/2}, \quad (5.7)\end{aligned}$$

and

$${}_{N+1}\langle 0|\Psi^\dagger(0,0) = \overline{(\Psi(0,0)|0\rangle_{N+1})}. \quad (5.8)$$

By using (3.10), one can expand the state (5.7) by $\Phi_{N;\lambda}(z)$ again. The result is

$$\begin{aligned}&\Psi(0,0)|0\rangle_{N+1} \\ &= \sum_\lambda \frac{t^{|\lambda|}(t^{-1})_\lambda^{(p,t)}}{h_{\lambda'}(t,p)}\Phi_{N;\lambda}(z)\prod_{j=1}^N\Big(\frac{(1/z_j;p)_\infty(tz_j;p)_\infty}{(t/z_j;p)_\infty(z_j;p)_\infty}\Big)^{1/2}. \quad (5.9)\end{aligned}$$

Then by (3.3), we obtain[15]

$$\begin{aligned}&\langle 0|\Psi^\dagger(\xi,\tau)\Psi(0,0)|0\rangle \\ &= \frac{A_N}{A_{N+1}}\sum_\lambda \frac{t^{2|\lambda|}\big((t^{-1})_\lambda^{(p,t)}\big)^2}{h_\lambda(p,t)h_{\lambda'}(t,p)}\mathcal{N}(\lambda)e^{-i(\mathcal{E}(\lambda)\tau-\mathcal{P}(\lambda)\xi)}. \quad (5.10)\end{aligned}$$

The factor $(t^{-1})_\lambda^{(p,t)}$ in (5.10) vanishes if the Young diagram $\mathcal{D}(\lambda)$ contains the lattice point $(s,r+1)$. This indicates that only the states which contains minimal $s-1-$quasi-particle and $r-$quasi-hole excitations can contribute to the intermediate state. Since $\Psi(0,0)$ annihilates one pseudo-particle from the $N+1-$particle ground state, this feature is again consistent with the fractional exclusion statistics with statistical paprmeter $g=r/s$ discussed in §4.

## 5.2 Thermodynamic limit

Let us next consider the thermodynamic limit $N, L \to \infty$, with $N/L = n$ fixed. Let us first consider the dynamical density-density correlation function. Noting the remark below (5.4) and using the formulae given in §A.1, we obtain the following expression.

$$\langle 0|\rho(\xi,\tau)\rho(0,0)|0\rangle = \mathcal{C}\Big(\prod_{i=1}^{s}\int_0^{\infty} d\alpha_i\Big)\Big(\prod_{j=1}^{r}\int_0^1 d\beta_j\Big) \operatorname{sh}^2\frac{\pi n}{c}(\sum_{i=1}^{s}\alpha_i + \sum_{j=1}^{r}\beta_j)$$
$$\times F(s,r;g|\{\alpha_i,\beta_j\})\cos(\mathcal{P}_{r,s}\xi)e^{-i\mathcal{E}_{r,s}\tau}. \quad (5.11)$$

Here we set $\alpha_i = \lambda_i/N$ and $\beta_j = \lambda'_j/N$. The quantities $\mathcal{P}_{r,s}$ and $\mathcal{E}_{r,s}$ are the thermodynamic limit of $P_N(\lambda)$ and $\mathcal{E}_N(\lambda)$, respectively:

$$\mathcal{P}_{k,l} = c\sum_{i=1}^{l}\rho_P(i) + c\sum_{j=1}^{k}\rho_H(j), \quad (5.12)$$

$$\mathcal{E}_{k,l} = c^2\sum_{i=1}^{l}\epsilon_P(i) + c^2\sum_{j=1}^{k}\epsilon_H(j) \quad (5.13)$$

with

$$\rho_P(i) = 2\operatorname{sh}\frac{\pi n\alpha_i}{c}\operatorname{ch}\frac{\pi n(\alpha_i+g)}{c}, \quad (5.14)$$

$$\rho_H(j) = \frac{2}{g}\operatorname{sh}\frac{\pi ng\beta_j}{c}\operatorname{ch}\frac{\pi ng(1-\beta_j)}{c}, \quad (5.15)$$

$$\epsilon_P(i) = 2\operatorname{sh}\frac{\pi n\alpha_i}{c}\operatorname{sh}\frac{\pi n(\alpha_i+g)}{c}, \quad (5.16)$$

$$\epsilon_H(j) = \frac{2}{g}\operatorname{sh}\frac{\pi ng\beta_j}{c}\operatorname{sh}\frac{\pi ng(1-\beta_j)}{c}. \quad (5.17)$$

Due to the argument below (5.4), $\rho_P$ and $\epsilon_P$ (resp. $\rho_H$ and $\epsilon_H$) are regarded as the quasi-particle (resp. quasi-hole) contribution to the excitation momentum and energy. The normalization constant $\mathcal{C}$ is obtained as

$$\mathcal{C} = 4\Big(\frac{\pi n}{c}\Big)^{r+s}g^{r(s-1)}r^2\Gamma^2(r)\mathcal{A}(s,r;g), \quad (5.18)$$

$$\mathcal{A}(l,k;g) = \frac{\Gamma^l(g)\Gamma^k(1/g)}{\prod_{i=1}^{l}\Gamma^2(k-g(i-1))\prod_{j=1}^{k}\Gamma^2(1-(j-1)/g)}. \quad (5.19)$$

The form factor $F(l,k,g|\{\rho_i,\tau_j\})$ is evaluated as

$$F(l,k;g|\{\alpha_i,\beta_j\})$$
$$= \prod_{j=1}^{k}\prod_{i=1}^{l}\text{sh}^{-2}\frac{\pi n}{c}(\alpha_i+g\beta_j)$$
$$\times \frac{\prod_{i<j}^{l}|\text{sh}\frac{\pi n}{c}(\alpha_i-\alpha_j)|^{2g}\prod_{i<j}^{k}|\text{sh}\frac{\pi ng}{c}(\beta_i-\beta_j)|^{2/g}}{\prod_{i=1}^{l}[\text{sh}\frac{\pi n}{c}\alpha_i\text{sh}\frac{\pi n}{c}(\alpha_i+g)]^{1-g}\prod_{j=1}^{k}[\text{sh}\frac{\pi ng}{c}\beta_j\text{sh}\frac{\pi ng}{c}(1-\beta_j)]^{1-1/g}}.$$
(5.20)

Similarly, by using the formulae in §A.2, the thermodynamic limit of the one-particle Green function is obtained as

$$\langle 0|\Psi^\dagger(\xi,\tau)\Psi(0,0)|0\rangle$$
$$= \mathcal{D}\Big(\prod_{i=1}^{s-1}\int_0^\infty d\alpha_i\Big)\Big(\prod_{j=1}^{r}\int_0^1 d\beta_j\Big)\exp\{\frac{2\pi n(1-g)}{c}(\sum_{i=1}^{s-1}\alpha_i+\sum_{j=1}^{r}\beta_j)\}$$
$$\times F(s-1,r;g|\{\alpha_i,\beta_j\})e^{-i(\mathcal{E}_{r,s-1}\tau-\mathcal{P}_{r,s-1}\xi)}, \quad (5.21)$$

where

$$\mathcal{D}=4g^{r(s-1)+g}\Big(\frac{\pi n}{c}\Big)^{r+s-1}(1-e^{-\frac{2\pi ng}{c}})^{g-1}e^{\frac{\pi ng(g-1)}{c}}\frac{\Gamma^2(r)}{\Gamma^2(g)}\mathcal{A}(s-1,r;g) \quad (5.22)$$

It should be remarked that in the non-relativistic limit $c\to\infty$, the results (5.11) and (5.21) coincide with those obtained by Ha in the CSM[10] up to scalar multiple.

## 5.3 Low energy asymptotic behavior

The exact results obtained in §5.2 allows us to analyze their low energy asymptotic behavior. Here 'low energy asymptotic behavior' means that we analytically continue the time variable $\tau$ to the imaginary time $-i\tau$ and consider the leading behavior for $\tau \gg 1$. The leading contribution is obtained from the saddle points, i.e. the points satisfying $\mathcal{E}_{r,s}=0$ (resp. $\mathcal{E}_{r,s-1}=0$) for the density-density correlation function (resp. for one-particle Green function). From (5.13), there are $r+1$ such points, where all of $\alpha_i$ and $r-m$ ($m=0,1,..,r$) of $\beta_j$ are 0 and remaining $m$ of $\beta_j$ are 1. Expanding around these points, we obtain

$$\mathcal{P}_{k,l}\xi \pm i\mathcal{E}_{k,l}\tau = 2\pi n\,\zeta_\mp\Big[\sum_{i=1}^{l}\alpha_i+\sum_{j=1}^{k-m}\beta_j\Big]$$

$$-2\pi n\zeta_\pm \sum_{a=1}^{m} \omega_a + 2mp_F\xi + \text{higer terms}, \qquad (5.23)$$

where we set $\zeta_\pm = \xi \text{ch}\frac{\pi g n}{c} \pm ic\tau \text{ sh}\frac{\pi g n}{c}$, $\omega_a = 1 - \beta_a$ and $p_F = \frac{c}{g}\text{sh}\frac{\pi g n}{c}$ denotes the fermi momentum. The expansion of the dynamical correlation functions (5.11) and (5.21) can be carried out in the same way as in the CSM[10]. We obtain their leading order behaviors as follows.

$$\langle 0|\rho(\xi,\tau)\rho(0,0)|0\rangle$$
$$\approx \tilde{\mathcal{C}}\, \mathcal{I}_1^g(1|s,r)\Big\{\Big(\frac{1}{\zeta_+}\Big)^2 + \Big(\frac{1}{\zeta_-}\Big)^2\Big\} + \sum_{m=1}^{r} \mathcal{C}_m \Big(\frac{1}{\zeta_+\zeta_-}\Big)^{m^2/g}\cos(p_F m\xi), \qquad (5.24)$$

$$\langle 0|\Psi^\dagger(\xi,\tau)\Psi(0,0)|0\rangle$$
$$\approx \sum_{m=0}^{r} \mathcal{D}_m \Big(\frac{1}{2\pi n\zeta_+}\Big)^{(m-g)^2/g} \Big(\frac{1}{2\pi n\zeta_-}\Big)^{m^2/g} \exp i(p_F m\xi + i\mu\tau). \qquad (5.25)$$

Here $\mu = c^2 \text{ch}\frac{\pi g n}{c}$ being the chemical potential and

$$\tilde{\mathcal{C}} = g^{r(s-1)}\mathcal{C}\Big(\frac{\pi n}{c}\Big)^{-r-s}\Big(\frac{1}{2\pi n}\Big)^2 \qquad (5.26)$$

$$\mathcal{C}_m = g^{-2ms}\tilde{\mathcal{C}}\binom{r}{m}\Big(\frac{1}{2\pi n}\Big)^{2+2m^2/g}\Big(\frac{\pi n g}{c}\Big/\text{sh}\frac{\pi n g}{c}\Big)^{2m^2/g}\mathcal{I}_1^g(1|s,r-m)\,\mathcal{I}_2^g(1|m), \qquad (5.27)$$

$$\mathcal{D}_m = g^{1-g+r(s-1)-2m(s-1)}\mathcal{D}\binom{r}{m}\Big(\frac{\pi n}{c}\Big)^{-r-s+1}\Big(\frac{\pi n g}{c}\Big/\text{sh}\frac{\pi n g}{c}\Big)^{g-1-2m+2m^2/g}$$
$$\times \tilde{\mathcal{I}}_1^g(1|s-1,r-m)\,\mathcal{I}_2^g(1|m) \qquad (5.28)$$

where

$$\mathcal{I}_1^g(z|l,k-m) = \Big(\prod_{i=1}^{l}\int_0^\infty d\alpha_i\Big)\Big(\prod_{j=1}^{k-m}\int_0^\infty d\beta_j\Big)\prod_{i=1}^{l}\prod_{j=1}^{k-m}(\alpha_i+g\beta_j)^{-2}$$
$$\times \frac{\prod_{i<j}|\alpha_j-\alpha_i|^{2g}\prod_{i<j}|\beta_j-\beta_i|^{2/g}}{\prod_{i=1}^{l}\alpha_i^{1-g}\prod_{j=1}^{k-m}\beta_j^{1-1/g}}$$
$$\times \exp\{-z(\sum_{i=1}^{l}\alpha_i + \sum_{j=1}^{k-m}\beta_j)\}\times \mathcal{I}_m(\alpha_i,\beta_j), \qquad (5.29)$$

$$\mathcal{I}_2^g(z|m) = \Big(\prod_{a=1}^{m}\int_0^\infty d\omega_a\Big)\frac{\prod_{a<b}|\omega_b-\omega_a|^{2/g}}{\prod_{a=1}^{m}\omega_a^{1-1/g}}\exp\{-z\sum_{a=1}^{m}\omega_a\}. \qquad (5.30)$$

In (5.29), the function $\mathcal{I}_m(\alpha_i, \beta_j)$

$$\mathcal{I}_m(\alpha_i, \beta_j) = \begin{cases} \left(\frac{\pi n}{c}\right)^2 (\sum_i^l \alpha_i + \sum_j^k \beta_j)^2 & \text{for } m = 0; \\ \text{sh}^2 \frac{\pi n m}{c} & \text{for } m \neq 0 \end{cases}.$$

The factor $\tilde{\mathcal{I}}_1^g(1|s-1, r-m)$ in (5.28) is obtained from $\mathcal{I}_1^g(1|l, k-m)$ by setting $l = s - 1$, $k = r$ and replacing $\mathcal{I}_m(\alpha_i, \beta_j)$ with

$$\begin{cases} \exp\left\{\frac{2\pi n}{c}(1-g)(\sum_i^{s-1} \alpha_i + \sum_j^r \beta_j)\right\} & \text{for } m = 0; \\ \exp\left\{\frac{2\pi n m}{c}(1-g)\right\} & \text{for } m \neq 0 \end{cases}.$$

In the derivation of (5.24) and (5.25), we carried out analytic continuation[31, 10]

$$\mathcal{I}_1^g(z|l, m) = \left(\frac{1}{z}\right)^{gl^2 - 2lm + 2\delta_{m,r} + m^2/g} \mathcal{I}_1^g(1|l, m), \tag{5.31}$$

$$\mathcal{I}_2^g(z|m) = \left(\frac{1}{z}\right)^{m^2/g} \mathcal{I}_2^g(1|m). \tag{5.32}$$

The asymptotic behaviors (5.24) and (5.25) coincides with those of the CSM[10].

## 6 Finite-size scaling

The exact energy spectrum obtained in §4 allows one to analyze the finite-size scaling of the model in the thermodynamic limit $N, L \to \infty$ and $N/L = n$ being fixed[32, 33, 34, 26, 35]. From this we identify the CFT describing the low-energy behavior of the model.

First of all, from (4.5) we obtain the finite-size correction to the ground state energy as

$$\lim E_N^{(0)} = L\varepsilon_0 - \frac{\pi v}{6L} g + O\left(\frac{1}{L^2}\right), \tag{6.1}$$

where we identified the ground state energy density with $\varepsilon_0 = \frac{c^3}{\pi g} \text{sh} \frac{\pi g n}{c}$ and the velocity of the elementary excitation with $v = c \, \text{sh} \frac{\pi g n}{c}$.

Comparing the result (6.1) with the general theory of the finite-size scaling[33, 34], one may suspect that the central charge is given by $g$. However this is not the correct identification[26]. The central charge should be identified with 1. This can be justified, for example, by calculating the low temperature expansion of the free energy from (4.9). Instead of doing this, we here justify it by deriving the whole conformal dimensions associated with the elementary excitations.

Let us consider the elementary excitations associated with the particle number change by $\Delta N$ and the transfer of the $\Delta D$ particles from the left Fermi point to the right one. We evaluate the following finite-size correction to the excitation energy and momentum.

$$\lim \Big[ c^2 \sum_{j=1}^{N+\Delta N} \text{ch}\frac{2\pi}{cL}\Big( g\Big(j - \frac{N+\Delta N + 1}{2}\Big) + \Delta D \Big) - E_N^{(0)} \Big]$$
$$= \mu \Delta N + \frac{2\pi v}{L}\Big[\frac{g}{4}\Delta N^2 + \frac{1}{g}\Delta D^2\Big] + O(\frac{1}{L^2}), \qquad (6.2)$$

$$\lim \ c \sum_{j=1}^{N+\Delta N} \text{sh}\frac{2\pi}{cL}\Big( g\Big(j - \frac{N+\Delta N + 1}{2}\Big) + \Delta D \Big)$$
$$= 2p_F \Delta D + \frac{2\pi \text{ch}\frac{\pi g n}{c}}{L}\Delta N \Delta D + O(\frac{1}{L^2}), \qquad (6.3)$$

where $\mu = c^2 \text{ch}\frac{\pi g n}{c}$ is the chemical potential. Adding the energy $\frac{2\pi v}{L}(N^+ + N^-)$ and the momentum $\frac{2\pi \text{ch}\frac{\pi g n}{c}}{L}(N^+ - N^-)$ of the quasi-particles and quasi-holes exciting near the Fermi surface, we obtain the finite-size corrections up to the order $1/L$:

$$\Delta E = \mu \Delta N + \frac{2\pi v}{L}\Big[\frac{g}{4}\Delta N^2 + \frac{1}{g}\Delta D^2 + N^+ + N^-\Big], \qquad (6.4)$$

$$\Delta P = 2\pi p_F \Delta D + \frac{2\pi \text{ch}\frac{\pi g n}{c}}{L}\Big[\Delta N \Delta D + N^+ - N^-\Big]. \qquad (6.5)$$

From these, we obtain the right and left conformal dimensions $h^\pm$ as[15]

$$h^\pm(\Delta N; \Delta D; N^\pm) = \frac{1}{2}\Big[\frac{\Delta N}{2R} \pm \Delta D R\Big]^2 + N^\pm, \qquad (6.6)$$

where we set $R = 1/\sqrt{g}$. The conformal dimensions with $N^\pm = 0$ are nothing but those of the $U(1)$−primary fields in the $C = 1$ Gaussian theory. We hence identify the desired CFT with the $C = 1$ Gaussian theory with compactification radius $1/\sqrt{g}$.

The standard theory of CFT then yields that correlation functions of any operator $\mathcal{O}$ can be expanded by those of the primary fields and the descendent fields as follows.

$$< \mathcal{O}(\xi,t)\mathcal{O}(0,0) > \approx \sum_{\substack{\Delta N, \Delta D \\ N_+, N_-}} \text{const.}\frac{e^{ip_F \Delta D \xi}}{(\zeta_+)^{2h^+}(\zeta_-)^{2h^-}} \qquad (6.7)$$

It is instructive to compare this with the exact results obtained in §5. In the case $\mathcal{O}(\xi,t) = \rho(\xi,t)$, we have no particle number changes so $\Delta N = 0$. Then the formula (6.7) with (6.6) and the identification $\Delta D = m$ agrees with the low-energy behavior of the dynamical density-density correlation function (5.24).

On the other hand in the case of the one-particle Green function, we have particle number change $\Delta N = 1$. In such a case, we has to consider the statistics of the pseudo-particle annihilated by $\Psi$. Such statistics is formulated as the boundary condition of the wave function[26, 35]. We assume that the wave function of the pseudo-particles with the rapidity (4.3) obey the periodic boundary condition

$$e^{i\theta_j L} = 1. \tag{6.8}$$

This yields the following selection rule

$$\Delta D = \frac{g}{2}\Delta N \qquad (\text{mod } 1). \tag{6.9}$$

Substituting (6.9) with $\Delta N = 1$ into (6.6), we obtain

$$h^+ = \frac{(\Delta D + g)^2}{2g}, \qquad h^- = \frac{\Delta D^2}{2g}. \tag{6.10}$$

This agrees with the exact result (5.25), if one sets $\Delta D = m$. We hence conclude that the low-energy effective theory of the TRSM (as well as CSM) is described by the $C = 1$ Gaussian theory with compactification radius $1/\sqrt{g}$ and selection rule (6.9).

The meaning of the selection rule (6.9) in the context of the Gaussian theory is as follows. According to the argument in [7], a pseudo-particle carries a flux $\pi g$. The flux twists the boundary condition of the boson field $\phi(\sigma,\tau)$ describing the Gaussian theory from $\phi(\sigma + 2\pi i,\tau) = \phi(\sigma,\tau) + 2\pi R m$, $m \in \mathbf{Z}$ to $\phi(\sigma + 2\pi i,\tau) = \phi(\sigma,\tau) + 2\pi R(m + \frac{g\Delta N}{2})$. This twisted Gaussian theory is similar to the low-energy effective theory discussed by Ha[10].

# 7 Conclusion and discussions

We have discussed a description of the excitations and a calculation of the dynamical correlation functions in the trigonometric Ruijsenaars-Schneider model. In the process, we have applied the theory of the Macdonald symmetric functions. We have shown that for the rational coupling constant $g$, the excitations of the model obey the fractional exclusion statistics with statistical parameter $g$. We also have obtained the one-parameter deformations of the dynamical correlation functions of the Calogero-Sutherland model. From the finite-size scaling analysis, we have also shown that the trigonometric Ruijsenaars-Schneider model is effectively described by the $C = 1$ Gaussian theory with $R = 1/\sqrt{g}$ in the low-energy region under a new selection rule for the quantum numbers labeling the critical exponents.

Recently, Gorsky and Nekrasov have shown the equivalence between the trigonometric Ruijsenaars-Schneider model and the G/G gauged WZW model with a Wilson line insertion[22]. This is a natural deformation of the equivalence between the Calogero-Sutherland model and the 2d Yang-Mills theory[36, 37]. To show this equivalence at the amplitude level should deepen the understanding of both the G/G gauged WZW model and the Ruijsenaars-Schneider model.

In addition, in the classical level, the hyperbolic Ruijsenaars-Schneider model is known to have a deep connection with the sine-Gordon theory[18, 19]. This connection should be examined at the quantum level.

Another possible problem is a connection to the random systems such as random matrix model, 2d disordered system in condensed matter physics and the quantum chaos. The latter systems indicate certain universal behavior and the Calogero-Sutherland model provide these systems with universal functions. The parallel structure of the Ruijsenaars-Schneider models to the Calogero-Sutherland seems to suggest the existence of some one-parameter deformation of the random systems.

We hope to return to these problems in future.

# 8 Acknowledgments

The author would like to thank Olivier Babelon, Denis Bernard, Peter Forrester, Michio Jimbo, Norio Kawakami, Tetsuji Miwa and Sung-Kil Yang for stimulating discussions. He also would like to thank J.F. van Diejen, Takahiro Fukui, Jyoichi Kaneko, Vadim Kuznetsov, Ryu Sasaki, Kimio Ueno and Takashi Yamamoto for useful conversations and Simon Ruijsenaars for communication. He is also grateful to Jens Petersen for reading the manuscript and Denis Bernard and Jean-Bernard Zuber for their kind hospitality during his stay at Saclay. This work is supported by the Yukawa

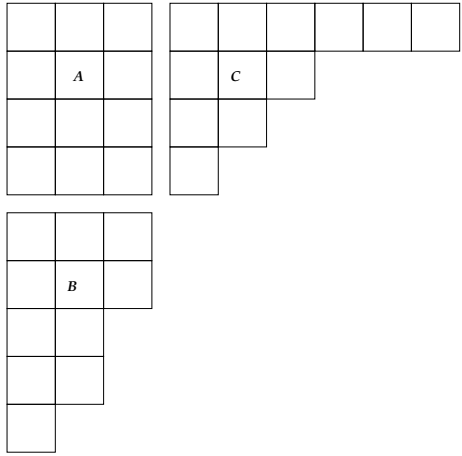

Figure 1: A Young diagram is divided into three sub-diagrams. This diagram describes the case $g = 3/4$.

memorial foundation.

# A  Formulae for the thermodynamic limit

## A.1  Dynamical density-density correlation function

Due to the remark below (5.4), only the partitions which Young diagram does not contain the cell $(s+1, r+1)$ can contribute to the density-density correlation function. Following Ha[10], we divide such Young diagrams into the following three sub-diagrams (Fig.1).

$$\begin{aligned}
\mathcal{A} &= \{(i,j) | 1 \leq i \leq s, 1 \leq j \leq r\}, \\
\mathcal{B} &= \{(i,j) | 1 \leq j \leq r, s+1 \leq i \leq \lambda'_j\}, \\
\mathcal{C} &= \{(i,j) | 1 \leq i \leq s, r+1 \leq j \leq \lambda_i\}.
\end{aligned} \quad (A.1)$$

Now let us first consider the factor $\mathcal{N}(\lambda)$ appearing in (5.4). According to the decomposition (A.1), $\mathcal{N}(\lambda)$ is decomposed into the following three factors.

$$\mathcal{N}(\mathcal{A}) = \prod_{j=1}^{r} \prod_{i=1}^{s} \frac{1 - p^{j-1} t^{N-i+1}}{1 - p^{j} t^{N-i}}, \quad (A.2)$$

$$\mathcal{N}(\mathcal{B}) = \prod_{j=1}^{r} \prod_{i=s+1}^{\lambda'_j} \frac{1-p^{j-1}t^{N-i+1}}{1-p^j t^{N-i}}$$

$$= \prod_{j=1}^{r} \frac{\Gamma_t(N-s+1+(j-1)/g)\Gamma_t(N-\lambda'_j+j/g)}{\Gamma_t(N-\lambda'_j+1+(j-1)/g)\Gamma_t(N-s+j/g)}, \quad (A.3)$$

$$\mathcal{N}(\mathcal{C}) = \prod_{i=1}^{s} \prod_{j=r+1}^{\lambda_i} \frac{1-p^{j-1}t^{N-i+1}}{1-p^j t^{N-i}}$$

$$= \prod_{i=1}^{s} \frac{\Gamma_p(\lambda_i+g(N-i+1))\Gamma_p(r+1+g(N-i))}{\Gamma_p(r+g(N-i+1))\Gamma_p(\lambda_i+1+g(N-i))}, \quad (A.4)$$

where

$$\Gamma_p(x) = \frac{(p;p)_\infty}{(p^x;p)_\infty}(1-p)^{1-x}. \quad (A.5)$$

In the thermodynamic limit $N, L \to \infty$ with $N/L = n$ fixed, we obtain

$$\mathcal{N}(\mathcal{A}) \to 1. \quad (A.6)$$

Noting $p^N \to e^{-\frac{2\pi n}{c}}$, $t^N \to e^{-\frac{2\pi n g}{c}}$ and the formula

$$\frac{\Gamma_t(z+a)}{\Gamma_t(z)} \to \left(\frac{1-t^z}{1-t}\right)^a \quad (A.7)$$

for $|z| \to \infty$, we obtained the thermodynamic limit of the remaining factors as follows.

$$\mathcal{N}(\mathcal{B}) \to \prod_{j=1}^{r} \left(\frac{1-e^{-\frac{2\pi n g(1-\beta_j)}{c}}}{1-e^{-\frac{2\pi n g}{c}}}\right)^{1/g-1}, \quad (A.8)$$

$$\mathcal{N}(\mathcal{C}) \to \prod_{i=1}^{s} \left(\frac{1-e^{-\frac{2\pi n(g+\alpha_i)}{c}}}{1-e^{-\frac{2\pi n g}{c}}}\right)^{g-1}. \quad (A.9)$$

Similarly, the factor $\chi^\lambda(p,t)$ is given by the product of the following three factors

$$\chi^{\mathcal{A}}(p,t) = \prod_{j=1}^{r} \prod_{i=1}^{s} (t^{i-1}-p^{j-1})$$

$$= (-)^{rs} p^{rs(r-1)/2}(1-t)^{r(s-1)} \frac{(p;p)_\infty}{(p^{r+1};p)_\infty}$$

$$\times \prod_{j=1}^{r} \frac{\Gamma_t(s-(j-1)/g)}{\Gamma_t(1-(j-1)/g)}, \tag{A.10}$$

$$\chi^{\mathcal{B}}(p,t) = \prod_{j=1}^{r} \prod_{i=s+1}^{\lambda'_j} (t^{i-1} - p^{j-1})$$

$$= (-)^{rs + \sum_j \lambda'_j} p^{-rs(r-1)/2 + \sum_{j=1}^{r}(j-1)\lambda_j} \prod_{j=1}^{r}(1-t)^{s-\lambda'_j}$$

$$\times \frac{\Gamma_t(\lambda'_j - (j-1)/g)}{\Gamma_t(s - (j-1)/g)}, \tag{A.11}$$

$$\chi^{\mathcal{C}}(p,t) = \prod_{i=1}^{s} \prod_{j=r+1}^{\lambda_i} (t^{i-1} - p^{j-1})$$

$$= t^{-rs(s-1)/2 + \sum_{i=1}^{s}(i-1)\lambda_i}$$

$$\times \prod_{i=1}^{s}(1-t)^{(\lambda_i - r)/g} \frac{\Gamma_p(\lambda_i - g(i-1))}{\Gamma_p(r - g(i-1))}. \tag{A.12}$$

Next we consider the factor $h_\lambda(p,t) = h_{\mathcal{A}}(p,t) h_{\mathcal{B}}(p,t) h_{\mathcal{C}}(p,t)$. The factor $h_{\mathcal{A}}(p,t)$ is evaluated as

$$h_{\mathcal{A}}(p.t) = \prod_{j=1}^{r} \prod_{i=1}^{s} (1 - p^{\lambda_i - j} t^{\lambda'_j - i + 1}). \tag{A.13}$$

In order to evaluate $h_{\mathcal{B}}(p,t)$, we decompose the diagram $\mathcal{B}$ further into the $r$-cells such that the $l$−th cell is given by $\{(i,j)|1 \leq j \leq l,\ \lambda'_{l+1}+1 \leq i \leq \lambda'_l\}$. If $\lambda'_l = \lambda'_{l+1}$, then the $l$−th cell is empty. We thus obtain the following expression.

$$h_{\mathcal{B}}(p,t) = \prod_{l=1}^{r} \prod_{j=1}^{l} \prod_{i=\lambda'_{l+1}+1}^{\lambda'_l} (1 - p^{\lambda_i - j} t^{\lambda'_j - i + 1})$$

$$= \prod_{j=1}^{r} (1-t)^{j(\lambda'_j - \lambda'_{j+1})} \Gamma_t(\lambda'_j + 1 - j/g)$$

$$\times \prod_{i<j}^{r} \frac{\Gamma_t(\lambda'_i - \lambda'_j + 1 + (j-i-1)/g)}{\Gamma_t(\lambda'_i - \lambda'_j + 1 + (j-i)/g)}. \tag{A.14}$$

Here note $\lambda'_{r+1} = s$. Similarly, we obtain

$$\begin{aligned}
h_{\mathcal{C}}(p,t) &= \prod_{l=1}^{s}\prod_{i=1}^{l}\prod_{j=\lambda_{l+1}+1}^{\lambda_l}(1-p^{\lambda_i-j}t^{\lambda'_j-i+1}) \qquad &\text{(A.15)}\\
&= \prod_{i=1}^{s}(1-p)^{i(\lambda_i-r)}\frac{\Gamma_p(\lambda_i-g(i-1))}{\Gamma_p(g)}\\
&\quad \times \prod_{i<j}^{s}\frac{\Gamma_p(\lambda_i-\lambda_j+g(j-i))}{\Gamma_p(\lambda_i-\lambda_j+g(j-i+1))}. &\text{(A.16)}
\end{aligned}$$

In the same way, we obtain

$$\begin{aligned}
h_{\mathcal{A}'}(t,p) &= \prod_{j=1}^{r}\prod_{i=1}^{s}(1-p^{\lambda_i-j+1}t^{\lambda'_j-i}) &\text{(A.17)}\\
h_{\mathcal{B}'}(t,p) &= \prod_{l=1}^{r}\prod_{j=1}^{l}\prod_{i=\lambda'_{l+1}+1}^{\lambda'_l}(1-p^{\lambda_i-j+1}t^{\lambda'_j-i})\\
&= \prod_{j=1}^{r}(1-t)^{j(\lambda'_j-\lambda'_{j+1})}\frac{\Gamma_t(\lambda'_j+1-j/g)}{\Gamma_t(1/g)}\\
&\quad \times \prod_{i<j}^{r}\frac{\Gamma_t(\lambda'_i-\lambda'_j+(j-i)/g)}{\Gamma_t(\lambda'_i-\lambda'_j+(j-i+1)/g)}, &\text{(A.18)}\\
h_{\mathcal{C}'}(t,p) &= \prod_{l=1}^{s}\prod_{i=1}^{l}\prod_{j=\lambda_{l+1}+1}^{\lambda_l}(1-p^{\lambda_i-j+1}t^{\lambda'_j-i})\\
&= \prod_{i=1}^{s}(1-p)^{i(\lambda_i-\lambda_{i+1})}\Gamma_p(\lambda_i+1-gi)\\
&\quad \times \prod_{i<j}^{s}\frac{\Gamma_p(\lambda_i-\lambda_j+1+g(j-i-1))}{\Gamma_p(\lambda_i-\lambda_j+1+g(j-i))}. &\text{(A.19)}
\end{aligned}$$

Combining (A.10) $\sim$ (A.19) and taking the thermodynamic limit, we obtain the following results

$$\begin{aligned}
\frac{(\chi^{\mathcal{A}}(p,t))^2}{h_{\mathcal{A}}(p,t)h_{\mathcal{A}'}(t,p)} &\to \left(\frac{2\pi}{Lc}\right)^{2rs}r^2 g^{2r(s-1)}\Gamma(r)^2\prod_{i=1}^{s}\prod_{j=1}^{r}(1-e^{-\frac{2\pi n(\alpha_i+g\beta_j)}{c}})^{-2}\\
&\quad \times \prod_{j=1}^{r}\frac{\Gamma^2(s-(j-1)/g)}{\Gamma^2(1-(j-1)/g)}, &\text{(A.20)}
\end{aligned}$$

$$\frac{(\chi^{\mathcal{B}}(p,t))^2}{h_{\mathcal{B}}(p,t)h_{\mathcal{B}'}(t,p)} \quad \rightarrow \quad \left(\frac{2\pi g}{Lc}\right)^{-r(s-1)} \prod_{j=1}^{r} \frac{\Gamma(1/g)}{\Gamma^2(s-(j-1)/g)}$$

$$\times \prod_{j=1}^{r} e^{\frac{-4\pi n(j-1)\beta_j}{c}} (1-e^{-\frac{2\pi n g \beta_j}{c}})^{1/g-1}$$

$$\times \prod_{i<j}^{r} (1-e^{-\frac{2\pi n g(\beta_i-\beta_j)}{c}})^{2/g}, \tag{A.21}$$

$$\frac{(\chi^{\mathcal{C}}(p,t))^2}{h_{\mathcal{C}}(p,t)h_{\mathcal{C}'}(t,p)} \quad \rightarrow \quad \left(\frac{2\pi}{Lc}\right)^{-s(r-1)} \prod_{i=1}^{s} \frac{\Gamma(g)}{\Gamma^2(r-g(i-1))}$$

$$\times \prod_{i=1}^{s} e^{\frac{-4\pi n(i-1)g\alpha_i}{c}} (1-e^{-\frac{2\pi n \alpha_i}{c}})^{g-1}$$

$$\times \prod_{i<j}^{s} (1-e^{-\frac{2\pi n(\alpha_i-\alpha_j)}{c}})^{2g}. \tag{A.22}$$

One should note that all the factors $1/L$ are absorbed into $n$ in (5.18) with $N$ appearing from the replacement $\sum_{\lambda_i} \rightarrow N \int d\alpha_i$ and $\sum_{\lambda'_j} \rightarrow N \int d\beta_j$.

## A.2  One-particle Green function

According to the same decomposition of the Young diagram, we obtain in the thermodynamic limit

$$\frac{((t^{-1})_{\mathcal{A}}^{(p,t)})^2}{h_{\mathcal{A}}(p,t)h_{\mathcal{A}'}(t,p)} \quad \rightarrow \quad \left(\frac{2\pi g}{Lc}\right)^{2r(s-1)} \prod_{i=1}^{s-1}\prod_{j=1}^{r} (1-e^{-\frac{2\pi n(\alpha_i+g\beta_j)}{c}})^{-2}$$

$$\times \prod_{j=1}^{r} \frac{\Gamma^2(s-(j-1)/g)}{\Gamma^2(1-(j-1)/g)} \tag{A.23}$$

$$\frac{((t^{-1})_{\mathcal{B}}^{(p,t)})^2}{h_{\mathcal{B}}(p,t)h_{\mathcal{B}'}(t,p)} \quad \rightarrow \quad \left(\frac{2\pi g}{Lc}\right)^{-r(s-1)} \prod_{j=1}^{r} \frac{\Gamma(1/g)}{\Gamma^2(s-(j-1)/g)}$$

$$\times \prod_{j=1}^{r} e^{\frac{-4\pi n(j-1-g)\beta_j}{c}} (1-e^{-\frac{2\pi n g \beta_j}{c}})^{1/g-1}$$

$$\times \prod_{i<j}^{r} (1-e^{-\frac{2\pi n g(\beta_i-\beta_j)}{c}})^{2/g}, \tag{A.24}$$

$$\frac{((t^{-1})_{\mathcal{C}}^{(p,t)})^2}{h_{\mathcal{C}}(p,t)h_{\mathcal{C}'}(t,p)} \quad \rightarrow \quad \left(\frac{2\pi}{Lc}\right)^{2r-s(r-1)-1-g} \prod_{i=1}^{s-1} \frac{\Gamma(g)}{\Gamma^2(r-gi)}$$

$$\times \prod_{i=1}^{s-1}(1-e^{-\frac{2\pi n\alpha_i}{c}})^{g-1}$$
$$\times \prod_{i<j}^{s-1}(1-e^{-\frac{2\pi n(\alpha_i-\alpha_j)}{c}})^{2g}. \qquad (A.25)$$